\documentclass[
nopreprint
]{jasatex}

\usepackage{dcolumn}

\usepackage{color}   
\usepackage{graphicx}
\usepackage{amsmath,amsfonts,amssymb}%

\newcommand{\beq}[1]{  \begin{equation} \label{#1} }  
\newcommand{\eeq}{     \end{equation}}  
\newcommand{\bal}[1]{\begin{align} \label{#1} }
\def\[#1\]{\begin{align}#1\end{align}}

\def\bd#1{\mbox{\boldmath$\displaystyle\mathbf{#1}$} }    
\def\dd{\operatorname{d}}
\def\e{\operatorname{e}}

\def\tr{\operatorname{tr}}
\def\sgn{\operatorname{sgn}}
\def\dbar#1{\Bar{\Bar{#1}}}

\begin{document} 


\title[Multiple scattering by cylinders]{Multiple scattering by cylinders immersed in fluid: \\ high order approximations for the effective wavenumbers}%

\author{Andrew N.  Norris}
\email{norris@rutgers.edu}
\affiliation{Mechanical \& Aerospace Engineering,   
  Rutgers University,  Piscataway,  NJ 08854, USA }

\author{Jean-Marc Conoir}
\affiliation{Institut Jean Le Rond d'Alembert,  
 UPMC Univ Paris 06, UMR 7190, F-75005 Paris, France}%
       
\date{\today}

\begin{abstract}
Acoustic wave propagation in a fluid with a random assortment of identical cylindrical scatterers is considered.  While the leading order correction to  the effective wavenumber of the  coherent wave is well established at dilute areal density ($n_0 $) of scatterers,  in this paper  the higher order dependence of the  coherent wavenumber on  $n_0$ is  developed   in  several directions.  Starting from  the quasi-crystalline approximation (QCA)   a consistent method is described for continuing the Linton and Martin formula, which is second order in  $n_0$, to higher orders.  Explicit formulas are provided for corrections to the effective wavenumber up to O$(n_0^4)$.  Then, using the QCA theory as a basis,  generalized self consistent schemes are developed and compared  with self consistent schemes using other dynamic effective medium theories.  It is shown that the Linton and Martin formula provides a closed self-consistent scheme, unlike other approaches.  

\end{abstract}

\pacs{43.35Bf, 43.20.Fn, 43.20.Hq}
\keywords{scattering, cylinders, effective medium}
 \maketitle

\section{Introduction}\label{sec0}

It is often assumed that multiple scattering by a dilute array of scatterers in a perfect fluid may be described by the propagation of a coherent wave representing the acoustic field averaged over all possible scatterer configurations. The coherent wave has a complex-valued wavenumber, $k_{eff}$, often called the effective wavenumber. Its imaginary part accounts for  loss due to scattering in all directions. 
Foldy first derived an   expression for the effective wavenumber  of point scatterers (isotropic scattering) \cite{Foldy45}.  Subsequently, Waterman and Truell considered finite size scatterers (non-isotropic scattering) \cite{Waterman61} and obtained a second order correction to  Foldy{'}s formula in terms of the scatterer density. The averaged exciting field in \cite{Waterman61} is obtained by replacing the spherical scatterers by point scatterers with the same angle-dependent far-field scattering amplitude as the actual finite size scatterers. 
For cylindrical scatterers, Waterman and Truell{'}s approach provides $k_{eff}=k_{WT}$
where \cite{Angel05}
\beq{intro1}
k_{WT}^{2}=\big[ k -\frac{2i n_{0}}{k }f(0) \big]^{2}-\big[\frac{2i n_{0}}{k }f(\pi)\big]^{2},
\eeq
with $f(\theta)$, defined in \S\ref{sec1}, the far-field scattered amplitude in direction $\theta$ for each cylinder and $n_{0}$ the number of scatterers per unit area.

In 1967  Lloyd and Berry \cite{Lloyd67} proposed an explicit expression for spherical scatterers different from that of Waterman and Truell. While  Lloyd and Berry{'}s formula had been obtained for spherical scatterers by using a different method \cite{Linton06}, Linton and Martin derived its counterpart for cylindrical scatterers in 2005 \cite{Linton05}.  Linton and Martin{'}s formula may be recovered from Fikioris and Waterman{'}s dispersion relation \cite{Fikioris64} by expanding it in powers of $n_{0}$, under the assumption that $n_{0}/k^{2} <<1$, and by letting the radius of exclusion in the \textit{hole correction} tends to zero. For cylindrical scatterers, Linton and Martin{'}s approach yields $k_{eff}=k_{LM}$, 
\beq{intro2} 
k_{LM}^{2}=k^{2}-4in_{0}f(0)+\frac{8 n_{0}^{2}}{ \pi k^{2}}
\int\limits_{0}^{\pi} \cot (\frac{\theta}{2}) \frac{\dd}{\dd\theta}[f(\theta)]^{2} \dd\theta.
\eeq
This formula is only valid for symmetric scattering functions satisfying $f(\theta)=f(-\theta)$ \cite{Conoir10}. Its generalization to arbitrary $f$ is obtained  by replacing $f^2(\theta)$ with 
$f(\theta)f(-\theta)$ in \eqref{intro2}, see below and \cite{Conoir10}.   
Linton and Martin{'}s formula is a consequence of the widely used  closure assumption known as the quasi-crystalline approximation (QCA).   
Martin  and Maurel \cite{Maurel08} demonstrated that QCA  agrees with the Lippmann-Schwinger approach (weak scattering) to O$(n_0^2)$.

The Independent Scattering Approximation (ISA) \cite{Lagendijk96}
is a  simplistic approximation in multiple-scattering theory, sometimes  used without justification.  It can be deduced directly from Eq. \eqref{intro1} and Eq. \eqref{intro2} by neglecting the term of second order in $n_{0}/k^{2}$, giving  $k_{eff}=k_{ISA}$ with
\beq{intro3}
k_{ISA}^{2}=k^{2}-4i n_{0}f(0).
\eeq


All of these methods are \textit{explicit} since $k_{eff}$ is given by a formula. 
But the question remains: which method is the most accurate?  This depends largely on the value of the concentration, $c=n_{0}\pi a^{2}$, 
where $a$ is the radius 
   for circular cylinders or  the radius of the enclosing cylinder for non-circular scatterers. 
It is evident that ISA, which is a method of order one in $n_{0}/k^{2}=c/\pi(k a)^{2}$, is the least precise. As for the two others, according to the analysis made by Linton and Martin in \cite{Linton05}, it is justified to think that $k_{LM}$ is more accurate than $k_{WT}$. However, the difference between $k_{WT}$ and $k_{LM}$ is small at low concentration, as shown in \cite{Derode06} when $c=6\%$.  Ref. \cite{Derode06} also  shows that the Waterman and Truell method or ISA fail while Linton and Martin{'}s 
gives   better results for higher concentration: $c=14\%$.

There are two other methods that deal with the far field scattered amplitudes, but they are \textit{implicit}; $k_{eff}$ is obtained by solving an equation. The most famous is the Coherent Potential Approximation (CPA), and its generalization (GCPA), both based on Dyson{'}s equation \cite{Sheng95}. There is also the Generalized Self Consistent Model (GSCM) which is derived  using a self consistent scheme applied to  Waterman and Truell{'}s formula \cite{Yang94}. The existence of a non trivial solution to the homogeneous system of linear equations obtained by Fikioris and Waterman also allows one to calculate a  $k_{eff}$  \cite{LeBas05}.
According to the finding of Cowan et al. \cite{Cowan98} the CPA is capable of accounting for  concentrations up to $60\%$ for glass beads in a liquid mixture of water and glycerol. Of course, scatterers do not radiate as much at this level of concentration as compared to those considered at the lower levels of concentration in \cite{Derode06}.  Nevertheless, the results of \cite{Cowan98} suggest that implicit methods can be more powerful than explicit ones even if, to our knowledge, no rigorous comparison has been made between the two types of methods.   Ref. \cite{Kim10} provides comprehensive numerical comparisons of eight different explicit and implicit methods. 

In this paper we search for explicit and implicit expressions of $k_{eff}$ that could allow us to consider higher concentrations for  multiple scattering problems.  For \textit{explicit methods}, the only way  to improve $k_{eff}$ is to extend  Linton and Martin{'}s formula to orders higher than 2 in concentration. This is accomplished in \S\ref{sec1}.  In this regard we note that  Waterman and Truell{'}s formula is of order 2 in concentration, and that it is is not possible to go further.  For \textit{implicit methods} we use the  self consistent scheme as used by Yang and Mal in \cite{Yang94}. When applied to the ISA  this self consistent scheme leads to CPA.  It is applied to the generalized Linton and Martin formula in \S\ref{sec2} where the physical meaning of the self consistent scheme is discussed.  In particular, we  obtain a new result that generalizes the CPA. In \S\ref{sec3} we compare all the effective wavenumbers in the Rayleigh limit (low frequencies).

\section{Higher order theory}\label{sec1}

\subsection{Multiple scattering formulation}\label{sec1.1}

The  problem is formulated  in terms of the pressure $\psi (\vec{r})$ which satisfies the Helmholtz equation in the interstitial space between scatterers: 
\beq{10}
 \nabla^2 \psi  + k^2 \psi = 0,
 \eeq 
 where $
k  =  {\omega}/{c} $ and $ c$ is the speed of sound.  Time harmonic dependence 
$e^{-i\omega t}$ is assumed. 
Consider first a system of $N$ scatterers with fixed positions, for which the 
 total field can be expressed
\beq{-4}
\psi  (\vec{r}) 
=\psi_{inc}(\vec{r}) + \sum_{j=1}^N T(\vec{r}_j)
{ {\psi_{E}(\vec{r},\vec{r}_j)} }.
\eeq
Here, $\psi_{inc}$ is the incident wave, $\psi_{E}(\vec{r}, \vec{r}_j)$ is the exciting field for scatterer $j$,  and  $T(\vec{r}_j)$ its scattering operator.  Focusing on scatterer $j=1$ as representative, we have 
\beq{-41}
{ {\psi_{E}(\vec{r},\vec{r}_1)} } =\psi_{inc}(\vec{r}) + \sum_{j=2}^N  T(\vec{r}_j)  \psi_{E}(\vec{r}, \vec{r}_j) .
\eeq
If, as is the case here,  the positions are not known, it becomes necessary to assume some type of  statistical description.   
The number density $n$ for $N$ discrete scatterers with centers at $\vec{r}_1$, $\vec{r}_2 \ldots$, is  ${ n(\vec{r}_1, \vec{r}_2
, \ldots , \vec{r}_N  ) } = N
p(\vec{r}_1, \vec{r}_2
, \ldots , \vec{r}_N  )
$, where $p$ is the probability density.   Conditional densities, defined by fixing the position of one scatterer, satisfy \cite{Waterman61} 
${ n( \vec{r}_2
, \ldots , \vec{r}_N | \vec{r}_1  ) } = (N-1)
p( \vec{r}_2
, \ldots , \vec{r}_N | \vec{r}_1 )
$.  For our purposes we need only the conditional number density of the $j^{th}$ scatterer if a scatterer is known to be at $ \vec{r}_1$: ${ n( \vec{r}_j\, 
 | \vec{r}_1  ) } = (N-1)
p( \vec{r}_j \, |
 \vec{r}_1 ),\, \, 
$ $j\ne 1$, satisfying $\int \dd \vec{r}_j\, { n( \vec{r}_j\, 
 | \vec{r}_1  ) } = N-1
$.
The analog of \eqref{-41} is  then an  equation for the configurationally averaged field,   
$\langle \psi_E (\vec{r}| \vec{r}_1)\rangle$, 
\bal{3}
&\langle \psi_E (\vec{r}| \vec{r}_1)\rangle =
 \psi_{inc} (\vec{r}) 
 \nonumber \\ & \quad
+ \int\dd 
 \vec{r}_j \,
{{ n(\vec{r}_j|  \vec{r}_1) }}
 T(\vec{r}_j) 
\langle \psi_E (\vec{r}| \vec{r}_j ,  \vec{r}_1 )\rangle .
\end{align}
As noted by \cite{Waterman61}, ``The fact that the exciting field with one scatterer fixed is given in terms of the field with two scatterers fixed is the basic difficulty in multiple scattering".   We adopt perhaps the simplest solution to this quandary, the QCA  (quasi-crystalline approximation), under which assumption \eqref{3} reduces to  
\bal{36-}
&\langle \psi_E (\vec{r}\, | \vec{r}_1)\rangle =
 \psi_{inc} (\vec{r}) 
  \nonumber \\ & \quad
  + \int\dd 
 \vec{r}_j \,
{{ n(\vec{r}_j | \vec{r}_1) }}
T(\vec{r}_j) 
\langle \psi_E (\vec{r}\, | \vec{r}_j)\rangle . 
\end{align}

The scattering operator for every scatterer is assumed to have translational invariance with
\bal{-13}
  T  (\vec{0} ) e^{ik x} &= 
  \sum\limits_n i^n T_n
 H^{(1)}_n(k r ) 
 e^{in \theta } 
   \nonumber \\ &
  \stackrel{\simeq}  {\text{\tiny{$(r\rightarrow \infty)$}}}
 \,  \sqrt{\frac{2}{\pi k  r} }
  e^{i( k  r - \frac\pi{4}) }\,
{  f  (\theta)} ,
\end{align}
and therefore  the Fourier series for $f$ is 
\beq{-14}
 {  f (\theta)  }= 
 \sum\limits_n T_ne^{in \theta}.
 \eeq
The assumed form of the operator in \eqref{-13} means that each scatterer 
has the same scattering behavior.  
We consider plane wave incidence, $
  \psi_{inc} (\vec{r})
 =  A  e^{ik x}$. 
The exciting potential $\langle\psi_E( \vec{r},\vec{r}_j )\rangle$ satisfies the Helmholtz equation and is regular function at the point $\vec{r_j}$, it can therefore be expressed
 \beq{-4-}
\langle \psi_E (\vec{r}| \vec{r}_j)\rangle 
 = \sum\limits_n 
{ A_n ( \vec{r}_j) }\, 
 J_n(k\rho_j )  
 e^{in \theta (\vec{\rho}_j)},  
 \eeq
 where $\vec{\rho}_j = \vec{r}- \vec{r}_j$, $\theta (\vec{\rho}_j)=\arg (\vec{\rho}_j)$. 
 Substituting from \eqref{-4-} into the configurational average equation \eqref{36-} and using the addition theorem for cylinder functions 
 \cite{Conoir10}, yields 
 \bal{-=9}
 & A_n(\vec{r}_1) = 
 i^n e^{ik x_1} + \sum\limits_p (-1)^p \, T_{n+p} 
 \int_{S^{+}}
 n(\vec{r}_j| \vec{r}_1)
 \nonumber \\ & 
 \quad \times    A_{n+p} ( \vec{r}_j) H^{(1)}_n(k\rho_j ) e^{in \theta (\vec{\rho}_j)}
  \dd 
 \vec{r}_j 
 ,
 \end{align}
 on the half-space  $S^{+}=\{ x>0\}$. 
The effective wavenumber $\xi$ defines the coherent wave according to the assumed functional form for each $A_n$: 
\beq{-6}
{ A_n(\vec{r}_j) }
=i^{n}{ A_n }e^{i\xi x_j}. 
\eeq
Use of the addition theorem for cylindrical functions then 
implies 
\beq{-5}
A_n e^{i{\xi}{x}_1} = A e^{ik x_1}
+\sum\limits_p   T_{n+p}A_{n+p}L_{p}(\xi) ,
\eeq
with 
\bal{-55} 
  L_{p}(\xi)&= i^{-p} \int_{S^{+}}\dd 
 \vec{r}_j \, n(\vec{r}_j| \vec{r}_1)
 \nonumber \\ & \qquad \times 
 H^{(1)}_p(k r_{j1} )
 e^{ip \theta (\vec{r}_{j1})} e^{i{\xi}{x}_{j}} ,
 \end{align}
 where $\vec{r}_{j1} = \vec{r}_{j}-\vec{r}_{1}$, 
 $\theta (\vec{r}_{j1}) = \arg(\vec{r}_{j1})$. 
 
 We assume a modified form of the hole correction with hole radius $b$,  
\beq{04=}
  n(\vec{r}_j| \vec{r}_1) = \begin{cases}
  n_0\big( 1+   h(r_{j1})\big), & r_{j1} >b, 
  \\
  0, & r_{j1} \le b, 
  \end{cases}
\eeq
where $h$ represents the deviation from the constant background value.  It satisfies
$h(r)\rightarrow 0$ as $r\rightarrow \infty$, with the stronger condition
\beq{-404}
\lim_{R\rightarrow \infty} R^{-2} \int_b^R h(r) r \dd r = 0 . 
\eeq
  Following, e.g.  \cite{Conoir10}, it may be shown that the integral \eqref{-55} reduces to 
\bal{-56} 
  L_{p}(\xi)&= 2\pi n_0  \bigg\{ \frac{ N_p(\xi b)}{\xi^2-k^2} + \frac{ M_p(\xi b)}{k^2}
  \bigg\}e^{i\xi x_1}  
  \nonumber \\  & \quad 
  + \frac{2 in_0 }{k(\xi -k)}e^{ik x_1}, 
   \end{align}
where
\begin{subequations}
\bal{jm25}
N_{p}(\xi b) &=\xi b J^{'}_{p}(\xi b)H^{(1)}_p (k b)
-k b J_{p} (\xi b) {H^{(1)}_p}^{'} (k b), 
\\
M_{p}(\xi b) & =    \int_b^\infty J_p(\xi r) H^{(1)}_p(k r ) h(r)  k^2 r \dd r .
\end{align}
\end{subequations}
The simple ``hole correction"   is $h=0$ and hence $M_p=0$; see Sec. IV.D of Ref. \cite{Linton05} for further discussion of the case $h\ne 0$.

\subsection{Matrix formulation}

Substituting from \eqref{-56} into \eqref{-5} and equating to zero the coefficients of $e^{i\zeta x_1}$ and 
$e^{ik x_1}$ yields two equations, known as the Lorentz-Lorenz law and the extinction theorem respectively, 
\begin{subequations}
\bal{-1}
 A_n + \frac{2\pi n_0}{k^2-\xi^2}
\sum\limits_{p=-\infty}^\infty {\cal N}_{n-p}(\xi b) T_pA_p  &= 0,
\\
A + \frac{2 in_0 }{k(\xi -k)} \sum\limits_{p=-\infty}^\infty T_pA_p &= 0,
\label{-1b}
\end{align}
\end{subequations}
where \eqref{-1} is satisfied by all $n\in \mathbb{Z}$ and 
\beq{-102}
{\cal N}_{p}(\xi b) =N_{p}(\xi b) + \big( \frac{\xi^2}{k^2}-1\big) M_{p}(\xi b).
\eeq
Equation \eqref{-1} is a homogeneous system which defines the effective wavenumber and the associated infinite eigenvector, while Eq.  \eqref{-1b} defines the amplitude of the eigenvector in terms of the excitation amplitude $A$.  
The identity $  {\cal N}_{-p}(\xi)=   {\cal N}_{p}(\xi)$ 
follows from the known properties of Bessel and Hankel functions.  
Equation \eqref{-1} with $h=0 \Rightarrow M_{p}=0$ is equivalent  to the system of equations studied by Linton-Martin \cite{Linton05} (their  equation (71)).    We focus on the solution of  Eq. \eqref{-1}. 


We introduce non-dimensional parameters
${y}$ and ${\epsilon } $ which  depend on $b$, along with some related vectors and matrices:
  \begin{subequations}\label{3-3}
\bal{3a}
y &= (\xi b)^2-(kb)^2 ,
\\
\epsilon &= -4in_0 b^2 
, \label{3b}
\\
{\bd b} &= {\bd T}^{1/2} {\bd a} ,\label{3d}  
\\
{\bd u} &= {\bd T}^{1/2} {\bd e} ,\label{3c}  
\\
{\bd Q} (y)  &=  (kb)^{-2}\, {\bd T}^{1/2}\bar{\bd Q}{\bd T}^{1/2} , \label{d}
\end{align}
\end{subequations}
where the  vectors ${\bd a}$ and ${\bd e}$, the  diagonal matrix ${\bd T}$, and the symmetric matrix 
$\bar{\bd Q}$,  are defined  
\begin{subequations}
\bal{3--3}
{\bd a} & = (\ldots, \, A_{-1} ,\, A_0  ,\, A_1  ,\ldots)^t,
\\
{\bd e} &= (\ldots, \, 1,\, 1,\, 1,\ldots)^t,
\\
T_{mn} &= T_n \delta_{mn} ,
\\
\bar{Q}_{mn}   & = 
\frac{ \frac{i\pi}{2} {\cal N}_{m-n}(\xi b)
- 1 } {   ( \xi/k)^2 - 1} , \label{3e}
\end{align}
\end{subequations}
Then Eq. \eqref{-1} can be expressed 
\beq{05-}
\big\{ y\big( {\bd I} - \epsilon {\bd Q} (y) \big) - 
 \epsilon {\bd u}{\bd u}^t \big\} {\bd b}
= 0.
\eeq
Setting the determinant of the matrix $\{ ..\}$ of this  infinite system of equations
yields the desired dispersion relation for $\xi$.  We next reduce this to a simple and transparent form, and obtain an expression  for the amplitude  vector ${\bd a}$. 

\subsection{Implicit solutions for $\xi$ and ${\bd a}$}

Multiply \eqref{05-} from the left by the inverse
of $\big(  {\bd I} -  \epsilon {\bd Q} \big) $, yielding
\beq{07-}
\big(   {\bd I} + {\bd w}{\bd u}^t \big) {\bd b}
= 0, 
\eeq
with infinite vector ${\bd w} =  - \epsilon y^{-1}  (  {\bd I} -  \epsilon {\bd Q}  )^{-1} {\bd u}$. 
Noting that  $\det ( {\bd I} + {\bd w}{\bd u}^t) = 1 + {\bd w}^t{\bd u}$, we deduce that 
 the  solution for $y$  can be expressed implicitly as 
\beq{334}
y =  \epsilon {\bd u}^t  \big[ {\bd I}- \epsilon   {\bd Q}(y )
\big]^{-1} {\bd u} .
\eeq

The dependence upon $b$ in Eq. \eqref{334} may be removed by introducing alternative  non-dimensional scalars $\bar{y}
={y}/{(kb)^2}$ and $\bar{\epsilon } ={\epsilon}/{(kb)^2}$, that is, 
\beq{+35} 
\bar{y} =\xi^2 k^{-2} - 1   ,
\quad
\bar{\epsilon }  = - 4 i n_0 k^{-2}  ,
\eeq
and writing $\bar{\bd Q}= \bar{\bd Q}(\bar{y} ) $, 
in terms of which Eq. \eqref{334}  becomes
\beq{036}
\bar{y} =    {\bd e}^t  
 \big[(\bar{\epsilon }  \,{\bd T})^{-1}- \bar{\bd Q}(\bar{y} ) 
\big]^{-1}  {\bd e} . 
\eeq
This formula clearly splits the dependence upon the scattering matrix $ {\bd T}$,  from that of  multiple interactions, $\bar{\bd Q}$. 
The choice of the symmetric matrix  $\bar{\bd Q} $, and hence ${\bd Q} $, is motivated by the observation that
at leading order the elements ${\cal N}_{n}$ are equal: ${\cal N}_{n}(k b) = 2/(i\pi)$ for all $n$.  Despite the apparent pole at $\xi = k$ in Eq. \eqref{3e},  the matrix  
$\bar{\bd Q} (\bar{y}) $ is a regular function of $\bar{y}$ at the origin since the limit and its derivatives exist as $\xi \rightarrow k$.   
The solution of Eq. \eqref{-1} may be expressed in implicit form 
through either of the identities   \eqref{334} or \eqref{036}.   We will find both useful in different circumstances.  The latter is simpler for considering general properties, while the former is useful for the particular limit as $b\rightarrow 0$.  

The identity $1 + {\bd w}^t{\bd u}=0$ also implies that the null vector of \eqref{07-} is of the form ${\bd b}= \alpha{\bd w}$ $(\alpha \ne 0)$.  The precise value of $\alpha$ follows from the extinction theorem \eqref{-1b}, and the vector of amplitudes $A_n$ 
can then be expressed as
\beq{0--3}
{\bd a}=\frac{2k}{k+\xi} 
\big[  \big( {\bd I} - \bar{\epsilon} \bar {\bd Q} (\bar{y}){\bd T} \big)  \big]^{-1}
{\bd e}.
\eeq

\subsection{Asymptotic expansion}

We seek an asymptotic expansion of $y=y_\epsilon $ in powers of the small parameter $\epsilon$:  
\beq{-2011}
y_\epsilon = \epsilon y_1 +  \epsilon^2 y_2 +  \epsilon^3 y_3 + \ldots . 
\eeq
The individual terms follow from \eqref{334} as 
\beq{-201}
y_n = \left . \frac1{n !} {\bd u}^t \frac{\dd^n }{\dd \epsilon^n}\bigg\{  
\epsilon   \big[ {\bd I}- \epsilon   {\bd Q}(y_\epsilon  )
\big]^{-1}  \bigg\} \right|_{ \epsilon =0}  {\bd u}.
\eeq

Using the expansion  $(1-x)^{-1} = 1+x+x^2 +\ldots$  in \eqref{-201} and noting that the derivative is evaluated at $\epsilon =0$ means that only a finite number of terms are necessary for a given $n$.  Thus, 
\bal{-202}
&y_n = \frac1{n !} {\bd u}^t \frac{\dd^n }{\dd \epsilon^n} \bigg\{ 
\epsilon^n   {\bd Q}^{n-1}(0  ) + 
\epsilon^{n-1}   {\bd Q}^{n-2}(\epsilon y_1  ) 
\nonumber \\ & 
  \ldots
 + \epsilon^2  {\bd Q} (\epsilon y_1 
 +\ldots + \epsilon^{n-2} y_{n-2}) 
\left . \bigg\}\right|_{ \epsilon =0}  {\bd u}.
\end{align}
We only need to expand each term in this finite series to obtain its O$(\epsilon^{n})$ contribution.  This is  simple for the first term $\epsilon^n   {\bd Q}^n(0  )$, 
and for the second it is $\epsilon^{n-1}   {\bd Q}^n(\epsilon y_1  ) 
\rightarrow \epsilon^n  y_1 
\frac{\dd }{\dd y}{\bd Q}^{n-2}(0)$ where $\frac{\dd }{\dd y}g(0)\equiv \frac{\dd }{\dd y}g(y)\big|_{ y =0}$.  Subsequent terms become more complicated but the procedure for finding their contribution is straightforward.  Thus, 
\bal{-203}
y_n &=   {\bd u}^t  \bigg\{ 
  {\bd Q}^{n-1}(0  ) + 
y_1  \frac{\dd }{\dd y}{\bd Q}^{n-2}(0)  
\nonumber \\ &  \qquad 
+
\frac{y_1^2}{2} \frac{\dd^2 }{\dd y^2}{\bd Q}^{n-3}(0) +
y_2  \frac{\dd }{\dd y}{\bd Q}^{n-3}(0)   
\nonumber \\ &  \qquad  
 \ldots
 + y_{n-2}  \frac{\dd }{\dd y}{\bd Q} (0)
\left . \bigg\}\right|_{ \epsilon =0}  {\bd u}.
\end{align}
  The following expansion includes all   terms up to fourth order in the small parameter 
$\epsilon$,
\bal{333}
y & = \epsilon {\bd u}^t{\bd u}
+\epsilon^2  {\bd u}^t{\bd Q}_0{\bd u}
\nonumber \\ & 
+\epsilon^3\big[ {\bd u}^t{\bd Q}_0^2{\bd u} +  ({\bd u}^t{\bd u}) {\bd u}^t{\bd Q}_0'{\bd u} 
\big] 
 \\
& 
+\epsilon^4
\big[{\bd u}^t{\bd Q}_0^3{\bd u} 
+
({\bd u}^t{\bd Q}_0{\bd u} ){\bd u}^t{\bd Q}_0'{\bd u}
 \nonumber \\
& 
+ 2({\bd u}^t {\bd u}) {\bd u}^t  {\bd Q}_0{\bd Q}_0'  {\bd u}
+\frac12 ({\bd u}^t {\bd u})^2  {\bd u}^t{\bd Q}_0''{\bd u}
\big]+\ldots ,
\nonumber
\end{align}
where 
\beq{305}
{\bd Q}_0 = {\bd Q}(0),\quad
{\bd Q}_0' = {\bd Q}'(0),\quad
{\bd Q}_0'' = {\bd Q}''(0) .
\eeq
The symmetry of ${\bd Q}^t = {\bd Q}$ has been used to simplify terms in the expansion \eqref{333}.

\subsection{Asymptotic expansion for finite $kb$}

Expanding the  function ${\cal N}_{p}(\xi b)$ for small $(\xi b - kb)$ yields
\beq{443}
\bar{Q}_{0mn} =  D^{(0)}_{m-n}(kb)  ,
\quad
\bar{Q}_{0mn}' =  D^{(1)}_{m-n}(kb)  ,
\eeq
\begin{subequations}
where
\bal{442}
&D_{p}^{(0)}(x) =  \frac{i\pi}{4}\bigg[(p^2-x^2)J_p(x)H_p^{(1)}(x) 
\nonumber \\ & \quad 
- x^2J_p'(x){H_p^{(1)}}'(x)
+ 2 M_p(kb)\bigg], 
\\
& D_{p}^{(1)}(x) =  - \frac12 D_{p}^{(0)}(x) 
+\frac{i\pi}{4} M_p(kb) 
\nonumber \\ &
\quad
+ \frac{1}{8}\big[p^2 - x^2
- i 2 \pi x^4 J_p(x)H_p^{(1)}(x) \big] .
\end{align}
\end{subequations}
Higher derivatives may  be found using the identity 
\bal{808}
&(\xi b)^2 {  N}_{p} '' + (\xi b)  {  N}_{p} ' + \big( 
(\xi b)^2 - p^2\big) {  N}_{p} 
\nonumber \\ & \qquad 
= - 2 (\xi b)^2  J_p(\xi b) H_p^{(1)}(kb) .
\end{align}

Using the above results, the expansion  to 
O$({\epsilon}^3)$  may be rewritten in terms of the areal density of scatterers, $n_0$, as 
\beq{543}
\xi^2 = k^2 + d_1 n_0+ d_2 n_0^2 + d_3 n_0^3 +  \ldots ,
\eeq
where 
\begin{subequations}\label{39}
\bal{+39a} 
d_1 &= -4 i \sum\limits_n T_n = - 4if(0), 
\\
d_2 & = -\frac{16}{ k^2}  \sum\limits_{m,n} D_{m-n}^{(0)}(kb) T_mT_n,\label{39b} 
\\
d_3 & = \frac{64i}{ k^4}  \sum\limits_{m,n,p} D_{m-n}^{(0)}(kb)D_{n-p}^{(0)}(kb) T_mT_nT_p 
\nonumber \\
&- \frac{16 d_1}{ k^4}\sum\limits_{m,n} D_{m-n}^{(1)}(kb) T_mT_n 
.
\end{align}
\end{subequations}
The dependence on $kb$ and the scattering function is contained in the coefficients $d_1$, $d_2$, etc. 

\subsection{Small $kb$ limit}


In this limit we derive  the terms in the series  
\beq{54}
\xi^2 = k^2 + \delta_1 n_0+ \delta_2 n_0^2 + \delta_3 n_0^3 + \delta_4 n_0^4 + \ldots .
\eeq
 It follows immediately from Eq. \eqref{39} that 
$\delta_1 = d_1$, while $\delta_2$ may be found  by letting $kb\rightarrow 0$ and using  
$D_{p}^{(0)}(x) = \frac{|p|}{2} + $O$(x)$ and 
$D_{p}^{(1)}(x) = \frac{p^2}{8} - D_{p}^{(0)}(x) + $O$(x)$ as $x\rightarrow 0$.   It is  easier, however, to begin with the small $kb$ expansion of $\bar{\bd Q}$. This allows us to deduce not only the terms in \eqref{543} to O$(n_0^3)$ but the next one.  Higher order terms can be found using the procedure described next. 
As $kb\rightarrow 0$ we have, using \eqref{jm25}, 
\beq{-34} 
{\cal N}_{p}(\xi b) \simeq  \frac{2}{i\pi}\bigg(\frac{\xi }{k}\bigg)^{|p|} . 
\eeq
Hence $\bar{\bd Q}$ becomes independent of $b$,
\beq{-35} 
\bar{Q}_{mn}(\bar{y}) \simeq    \big( (1+\bar{y})^{\frac{|m-n|}{2}} -1 \big) /\bar{y} ,
\eeq
from which  it follows that 
\bal{=12}
\bar{Q}_{0mn} &= \frac12 |m-n|,
\nonumber \\
\bar{Q}_{0mn} '&= \frac18 |m-n| (|m-n| - 2),
\\
\bar{Q}_{0mn} ''&= \frac1{24} |m-n| (|m-n| - 2)(|m-n| - 4), 
\nonumber  
\end{align}
etc. 
Now express the asymptotic expansion \eqref{333} in terms of the nondimensional parameters 
$\bar{y}$ and $  \bar{\epsilon }$ which  do not depend upon $b$,   
\bal{=33}
&\bar{y} = \bar{\epsilon} \tr {\bd T} 
+\bar{\epsilon}^2  {\bd e}^t {\bd T} \bar{\bd Q}_0{\bd T}{\bd e}
\nonumber \\ &
+\bar{\epsilon}^3\big[ 
 {\bd e}^t {\bd T} (\bar{\bd Q}_0{\bd T})^2  {\bd e}
 +  (\tr {\bd T}  ) {\bd e}^t {\bd T} \bar{\bd Q}_0'{\bd T} {\bd e}
\big] 
\nonumber \\ &
+\bar{\epsilon}^4
\big[
 {\bd e}^t {\bd T} (\bar{\bd Q}_0{\bd T})^3  {\bd e}
  +
({\bd e}^t {\bd T} \bar{\bd Q}_0{\bd T}{\bd e} )
{\bd e}^t {\bd T} \bar{\bd Q}_0'{\bd T}{\bd e}
\nonumber \\ & 
+ 2( \tr {\bd T})  
{\bd e}^t {\bd T} \bar{\bd Q}_0{\bd T}  \bar{\bd Q}_0'{\bd T}{\bd e}
+\frac12 (\tr {\bd T})^2  {\bd e}^t {\bd T} \bar{\bd Q}_0''{\bd T}{\bd e}
\big]
\nonumber \\ & 
+ \text{O} \big( \bar{\epsilon}^5 \big) . 
\end{align}
The matrices  $\bar{\bd Q}_0$, $\bar{\bd Q}_0'$, etc. are defined in the same way as in 
\eqref{305} for the matrix $\bar{\bd Q}(\bar{y})$ of \eqref{3e}. 
The coefficients in  Eq. \eqref{54} then  follow from eqs. \eqref{=12} and \eqref{=33} as 
\begin{subequations}\label{+33}
\bal{+33a} 
&\delta_1 = -4 i \sum\limits_n T_n ,
\\
&\delta_2   = -\frac{8}{ k^2}  \sum\limits_{m,n} |m-n|T_mT_n,
\\
&\delta_3  = \frac{16i}{ k^4} \sum\limits_{m,n,p} |m-n||n-p| T_mT_nT_p 
\nonumber \\ & \qquad
- \frac{2\delta_1}{ k^4}\sum\limits_{m,n} (m-n)^2T_mT_n 
- \frac{\delta_1\delta_2}{ 2k^2} , 
\label{+33b}
\\
&\delta_4   = \frac{32}{ k^6} \sum\limits_{m,n,p,q} |m-n||n-p||p-q|T_mT_nT_pT_q
\nonumber \\ & 
\qquad 
+ \frac{i8\delta_1}{ k^6}\sum\limits_{m,n,p} |m-n|(n-p)^2T_mT_nT_p
\nonumber \\
& - \sum\limits_{m,n } \bigg[ \frac{\delta_1^2}{ 3k^6} |m-n|^3  
+ \frac{2\delta_2}{ k^4}  (m-n)^2 \bigg] T_mT_n 
\nonumber \\  & 
\qquad 
- \frac{\delta_1^2\delta_2}{ 6k^4} 
- \frac{\delta_2^2}{ 2k^2} 
- \frac{\delta_1\delta_3}{  k^2} .
\label{+33c}
\end{align}
\end{subequations}

\subsection{The Linton-Martin formula generalized}

The coefficients in \eqref{+33} depend upon the far-field scattering function through its Fourier coefficients.  We now show that the coefficients can be expressed in terms of the function $f$ itself rather than its Fourier series.  Thus, 
\begin{subequations}\label{-33}
\bal{-33a} 
&\delta_1 =  - 4if(0), 
\\
& \delta_2  = \frac{8}{\pi k^2}
\int_0^\pi  \dd \theta  \cot (\frac\theta 2) 
\frac{\dd }{\dd \theta} [f(\theta) f(-\theta)] ,
\end{align}
\begin{align}
&\delta_3  = \frac{16i}{ \pi^2 k^4} 
\int_0^\pi \dd \theta \cot\frac\theta 2 \int_0^\pi \dd \bar \theta \cot\frac{\bar\theta}2 \, 
\mathbb{S} (\theta,\bar\theta)
\nonumber \\  & 
+ \frac{2\delta_1}{  k^4}  \frac{\dd^2 }{\dd \theta^2}  \big[f(\theta) f(-\theta)\big]_{\theta = 0}
 - \frac{\delta_1\delta_2}{ 2k^2}  ,
 \label{-33b} 
\\ 
&\delta_4   = -\frac{32}{\pi^3 k^6} 
\int\limits_0^\pi \dd \theta \cot\frac\theta 2 
\int\limits_0^\pi \dd \bar \theta \cot\frac{\bar\theta}2 
\int\limits_0^\pi \dd \dbar \theta \cot\frac{\dbar\theta}2 \, 
\nonumber \\ & \times
\mathbb{T} (\theta,\bar\theta,\dbar\theta)
- \frac{i8\delta_1}{\pi k^6} \int_0^\pi \dd \theta \cot (\frac\theta 2) 
\left. \frac{\partial }{\partial \bar\theta} \mathbb{S} (\theta,\bar\theta) \right|_{\bar\theta = 0}
\nonumber \\
&  - \frac{\delta_1^2}{ 3\pi k^6}
\int_0^\pi  \dd \theta  \cot (\frac\theta 2) 
\frac{\dd^3 }{\dd \theta^3} [f(\theta) f(-\theta)]
\nonumber \\ &
+\frac{2\delta_2}{ k^4}   \frac{\dd^2 }{\dd \theta^2}  \big[f(\theta) f(-\theta)\big]_{\theta = 0}
- \frac{\delta_1^2\delta_2}{ 6k^4} 
- \frac{\delta_2^2}{ 2k^2} 
- \frac{\delta_1\delta_3}{  k^2} ,
 \label{-33c} 
\end{align}
\end{subequations}
where 
\bal{-=5}
&\mathbb{S} (\theta,\bar\theta)  
= \frac14  \frac{\partial^2 }{\partial\theta\partial\bar\theta} \big\{     
      \nonumber  \\ &  \qquad \quad f(\theta) 
       \big[ f(-\bar\theta)f(\bar\theta-\theta)- f(\bar\theta)f(-\bar\theta-\theta) \big]
    \nonumber \\ & \qquad 
 +  f(-\theta) \big[ f(\bar\theta)f(\theta-\bar\theta)- f(-\bar\theta)f(\theta+\bar\theta) \big]
\big\}, 
\nonumber 
\\
&\mathbb{T} (\theta,\bar\theta,\dbar\theta)
= \frac18 \frac{\partial^3 }{\partial\theta\partial\bar\theta\partial\dbar\theta} 
\bigg\{  
 \nonumber \\ &    \quad
 f(\theta) f(\bar\theta-\theta)\big[ 
 f(-\dbar\theta) f(\dbar\theta-\bar\theta) -  f(\dbar\theta) f(-\dbar\theta-\bar\theta)
 \big]
\nonumber \\ 
& + 
 f(\theta) f(-\bar\theta-\theta)
\big[ 
 f(\dbar\theta) f(\bar\theta-\dbar\theta) -  f(-\dbar\theta) f(\bar\theta+\dbar\theta)
 \big]
\nonumber \\ 
& + 
 f(-\theta) f(\bar\theta+\theta)
 \big[ 
 f(\dbar\theta) f(-\bar\theta-\dbar\theta) -  f(-\dbar\theta) f(\dbar\theta-\bar\theta)
 \big]
\nonumber \\ 
& + 
 f(-\theta) f(\theta-\bar\theta)
 \big[ 
 f(-\dbar\theta) f(\bar\theta+\dbar\theta) -  f(\dbar\theta) f(\bar\theta-\dbar\theta)
 \big]\bigg\}.
 \nonumber 
\end{align}
We now justify these expressions. 

The first identity for $\delta_1$ in \eqref{-33} is obvious from the definition of   Eq. \eqref{-14}. 
Regarding $\delta_2$, we note that the product of $f(\theta)$ and $f(-\theta)$ may be written
\beq{321}
f(\theta) f(-\theta) =   
\sum\limits_{n=-\infty}^{\infty}
\sum\limits_{s=-\infty}^{\infty} T_n T_s e^{i(n-s)\theta}. 
\eeq
Interchanging the indices, implies that the double sum is   
\beq{5}
f(\theta) f(-\theta) =   \sum\limits_{n=-\infty}^{\infty}
\sum\limits_{s=-\infty}^{\infty} T_n T_s \cos (n-s) \theta. 
\eeq
This   identity is equivalent to  Eq. (83) of \cite{Linton05}
but without the restriction   $T_n = T_{-n}$ that was assumed there.   
The formula for $\delta_2$ follows from the relation \cite{Gradshteyn00} (Eq. 3.612(7))
\beq{3-2}
\frac1\pi \int_0^\pi \dd \theta \cot\frac\theta 2 \sin m \theta = \sgn (m).
\eeq
Turning to the coefficient $\delta_3$ in \eqref{+33b},  define
\bal{5=3}
S(\theta, \bar{\theta}) 
&= - \sum\limits_m\sum\limits_n\sum\limits_p (m-n)(n-p) 
\nonumber \\ & \qquad \times
T_mT_nT_p \, e^{i(m-n)\theta}e^{i(n-p)\bar\theta}
\nonumber \\
& = \frac{\partial^2 }{\partial\theta\partial\bar\theta} [f(\theta) f(\bar\theta-\theta)f(-\bar\theta)]. 
\end{align}
The individual terms in the triple sum are of the form $e^{i \cdots \theta}$, but an identical sum with the preferred dependence $\sin  \cdots \theta$ in each term can be obtained by an appropriate permutation of 
$2^2=4$ expressions.   The correct combination is 
\beq{08}
\frac14\big\{ 
S(\theta, \bar{\theta})  - S(\theta, -\bar{\theta})  - S(-\theta, \bar{\theta})  + S(-\theta, -\bar{\theta}) 
\big\}, \nonumber 
\eeq
which reduces to $\mathbb{S} (\theta,\bar\theta)$ as given. 
It may then be deduced, again using the identity \eqref{3-2}, that 
\bal{-562}
&\sum\limits_{m,n,p} |m-n||n-p| T_mT_nT_p 
\nonumber \\ & \quad 
= 
\frac1{\pi^2} \int_0^\pi \dd \theta \cot\frac\theta 2 \int_0^\pi \dd \bar \theta \cot\frac{\bar\theta}2 \, 
\mathbb{S} (\theta,\bar\theta) .
\end{align}
The second  term in \eqref{-33b}  follows directly from \eqref{5}. 
Regarding the coefficient $\delta_4$, the third term in the right member of  \eqref{-33c} follows in the same manner as  the integral for $\delta_2$, based on Eq. \eqref{5}. The second term in    \eqref{-33c}
may be deduced using the relation  
\bal{5=34}
& \frac{\partial }{\partial\bar\theta}\mathbb{S}(\theta, \bar{\theta}) 
= - \sum\limits_m\sum\limits_n\sum\limits_p (m-n)(n-p)^2 
\nonumber \\ &\qquad   \times 
T_mT_nT_p \, \sin (m-n)\theta  \cos(n-p)\bar\theta  
,
\end{align}
evaluated at $\bar\theta =0$ in combination with the  integral identity \eqref{3-2}.   Note that 
\begin{align}
& \left. \frac{\partial }{\partial \bar\theta} \mathbb{S} (\theta,\bar\theta) \right|_{\bar\theta = 0}
=  f''(0) \frac{\dd}{\dd \theta}[f(\theta)f(-\theta)] 
\nonumber \\ & 
-f'(0) \frac{\dd}{\dd \theta} [f(\theta)f'(-\theta)+f(-\theta)f'(\theta)]
\nonumber \\ &  
+ f(0) \frac{\dd}{\dd \theta}
[ f'(\theta)f'(-\theta)]
+ \frac{f(0)}2 \frac{\dd^3}{\dd \theta^3}[f(\theta)f(-\theta)] 
.\nonumber
\end{align}
Finally, the first term in the right member of  \eqref{-33c} may be obtained using the same type of argument used for the identity \eqref{-562}.  The  function analogous to $S$ now has three arguments, 
\bal{5=8}
&T(\theta, \bar\theta, \dbar\theta) 
= - i\sum\limits_m\sum\limits_n\sum\limits_p\sum\limits_q (m-n)(n-p) 
\nonumber \\ & \times 
(p-q)
T_mT_nT_p T_q\, e^{i(m-n)\theta}e^{i(n-p)\bar\theta}e^{i(p-q)\dbar\theta}
\nonumber \\
& \quad = \frac{\partial^3 }{\partial\theta\partial\bar\theta\partial\dbar\theta} [f(\theta) 
f(\bar\theta-\theta) f(\dbar\theta-\bar\theta)f(-\dbar\theta)]. 
\end{align}
and 
$2^3 = 8$ permutations are required in order to arrive at the correct quadruple summation, yielding $\mathbb{T} (\theta,\bar\theta,\dbar\theta)$.


\section{Simplifications based on the finite rank of ${\bd Q}$}\label{sec1.5}

The infinite matrix ${\bd Q}$ is in practice well approximated by a matrix of finite rank.  
This follows from the fact that the far-field scattering function $f(\theta)$ 
is an  entire function of the angular argument $\theta$ considered as a complex variable \cite{Muller55}. 
The scattering operator (matrix) ${\bd T}$ is therefore compact and has only a finite number of eigenvalues of finite size.  
At low frequency only the first few elements $T_m$ for $m$ near zero are important (monopole, dipole, etc.).   Furthermore, as we  will see in this section,   
${\bd Q}$ is of rank 2 in the high frequency limit.   Therefore, at any finite frequency the infinite system of equations is really not so in practice, and may be replaced by a finite system.   We first develop the solution for finite rank $n$, and then apply it to two important cases; the low frequency Rayleigh limit ($n=3$) and the high frequency limit ($n=2$). 
For the remainder of the paper we take $h$, introduced  in Eq. \eqref{04=}, to be zero, so that $M_p = 0$.


\subsection{${\bd Q}$ is of rank $n$}

As noted above  only a finite number of the elements $T_n$ are significant at any given frequency. If only $n$ are non-zero, then ${\bd Q}$ is of rank $n$. The matrix 
satisfies an homogeneous  equation of degree $n+1$,  
\beq{-43}
{\bd Q}^{n+1} +b_n {\bd Q}^{n} +b_{n-1} {\bd Q}^{n-1} + \ldots + b_1 {\bd Q} = 0 , 
\eeq
with $b_1 =(-1)^n (\det {\bd Q})$, $\cdots$, $b_n = -(\tr {\bd Q} )$,  
and therefore, 
\beq{-45}
( {\bd I} -   \epsilon {\bd Q} )^{-1}  = {\bd I} + 
\beta_1  \epsilon {\bd Q}  + \beta_2  \epsilon^2 {\bd Q}^2 + \ldots + \beta_n  \epsilon^n {\bd Q}^n,
\nonumber 
\eeq
where \beq{-46} 
\beta_j = \frac{ 1+ \sum_{k=j+1}^nb_k \epsilon^{n+1-k}}{ 1+ \sum_{m=1}^nb_m \epsilon^{n+1-m}}.
\eeq

Thus, using the fundamental result \eqref{334} it follows that $y= (\xi^2-k^2)b^2$ is given by 
\beq{-413}
y =  \epsilon {\bd u}^t {\bd u} + 
\sum\limits_{j=1}^n \epsilon^{j+1} \beta_j {\bd u}^t {\bd Q}^j {\bd u} .
\eeq
Since ${\bd Q}= {\bd Q}(y)$, \eqref{-413}  is an implicit equation for $y$ which could be solved by iteration, for instance.    It may also be written as 
\bal{-4135}
\xi^2 &= k^2 -4in_0f(0)  
\nonumber \\ & \quad 
- 16\frac{n_0^2}{k^2}  
\sum\limits_{j=1}^n {\bar{\epsilon}}^{j-1} \beta_j {\bd e}^t \big( {\bd T}\bar{\bd Q}\big)^j {\bd T}{\bd e} .
\end{align}

\subsubsection{Rayleigh limit, ${\bd Q}$ is of rank $3$} 
An important case is  $n=3$ which is useful at low frequency (Rayleigh limit) when the scattering matrix is well approximated by only three terms:
$T_0$ and $T_{\pm 1}$. For  $n=3$, 
\beq{-47}
{\bd Q}^{4}  - \text{I}_Q  {\bd Q}^{3}  +  \text{I}\!  \text{I}_Q{\bd Q}^{2} - \text{I}\! \text{I}\! \text{I}_Q {\bd Q} = 0 , 
\eeq
with
$  \text{I}_Q = \tr {\bd Q}$, 
 $   \text{I}\!  \text{I}_Q =\frac{1}{2}\big[ \big(\tr {\bd Q}\big)^{2} - \tr {\bd Q}^{2} \big]   $,
 $\text{I}\! \text{I}\! \text{I}_Q = \det {\bd Q}$. 
In this case 
the implicit equation for $y$  becomes 
\bal{-414}
y &=  \epsilon {\bd u}^t {\bd u} + 
\epsilon^2  \big[ 
\big( 1 - \epsilon \text{I}_Q+ \varepsilon^{2}   \text{I}\!  \text{I}_Q \big)\, {\bd u}^t {\bd Q}{\bd u}
\nonumber \\ & \quad
+  (1-\varepsilon \text{I}_Q)\epsilon {\bd u}^t  {\bd Q}^2 {\bd u} + \epsilon^{2} {\bd u}^t  {\bd Q}^3 {\bd u} \big] 
 \nonumber \\ & \quad \times 
\big[{ 1-  \epsilon \text{I}_Q  + \varepsilon^{2}  \text{I}\!  \text{I}_Q - \epsilon^3   \text{I}\! \text{I}\! \text{I}_Q}\big]^{-1}.
\end{align}
More detailed  results for the Rayleigh limit are presented in \S\ref{sec3} for the particular case of circularly cylindrical scatterers.

\subsection{High frequency limit}

In the high frequency limit $kb, \xi b \gg 1$, we have, from \eqref{jm25}, 
\beq{1001}
\frac{i\pi}2  N_{p}(\xi b)  \, \stackrel{=}  {\text{\tiny{$(\xi b\rightarrow \infty)$}}}
\, 
A+(-1)^p B, 
\eeq
with 
\beq{-3-3}
A= \frac{ (k+\xi)}{2\sqrt{ k\xi }} \,  e^{i(k-\xi) b} ,
\, \, 
B= \frac{ (k-\xi)}{i2\sqrt{ k\xi }} \,  e^{i(k+\xi) b} .
\nonumber
\eeq
The high frequency limit for elastic waves was discussed in \cite{Conoir10}, and the same methods developed there could be used for the acoustic problem.  It is instructive to note that \eqref{1001} and \eqref{d}, combined with $M_p=0$, implies that 
\beq{003}
 {\bd Q} = \alpha {\bd u}{\bd u}^t + \beta {\bd v}{\bd v}^t,  
\eeq
where 
 	\begin{align}
 &\alpha = ( A - 1)/y, 
 \qquad
 \beta = { B }/{ y}, 
 \nonumber \\ 
 &{\bd v} = {\bd T}^{1/2} (\ldots, \, -1,\, 1,\, -1,,\, 1, \ldots)^t,
\nonumber
 	\end{align}
 	with $ v_0 = T_0^{1/2}$. 
The matrix  ${\bd Q}$ is therefore   rank 2, and the wavenumber $\xi$ can    
be found using the methods described above.  

Equation \eqref{-414} reduces for rank 2 to 
\beq{-4143}
y =  \epsilon {\bd u}^t {\bd u} + 
\epsilon^2 \frac{
\big( 1 - \epsilon \tr {\bd Q}\big)\, {\bd u}^t {\bd Q}{\bd u}+  \epsilon {\bd u}^t  {\bd Q}^2 {\bd u}  }
{ 1-  \epsilon  \tr {\bd Q}  + \varepsilon^{2}   \det {\bd Q}}.
\eeq
Noting that  
\beq{004}
 {\bd u}^t{\bd u}=  {\bd v}^t{\bd v} = f(0), 
\quad
{\bd u}^t{\bd v}=  {\bd v}^t{\bd u} = f(\pi),
\eeq
and hence 
 	\begin{align}
\tr {\bd Q} &= (\alpha + \beta ) f(0), 
\nonumber \\
 \det {\bd Q}  &= \alpha  \beta \big( f^2(0) - f^2(\pi)\big), 
\nonumber
 	\end{align}
the equation for $\xi$ becomes
 	\begin{align}
& {y} = {\epsilon }f(0)
+  {\epsilon }^2 \times 
\nonumber \\ & 
  \big\{ \alpha f^2(0) + \beta f^2(\pi)
  	-  {\epsilon } f(0) \alpha   \beta  
 	(f^2(0) - f^2(\pi) )
 \big\}
\nonumber \\ &
 / 
\big\{ 1 -  {\epsilon } f(0) (\alpha + \beta )
 + {\epsilon }^2 \alpha   \beta \big( f^2(0) - f^2(\pi)\big)\big\}
 	 	.\nonumber 
 	\end{align}
 	This  simplifies  to   
\bal{009}
& (\xi^2 - k^2)^2 + 4in_0 (\xi^2 - k^2)(A+B) f(0) 
\nonumber \\ & \quad
- 16n_0^2 AB \big( f^2(0) - f^2(\pi)\big) =0, 
 	\end{align}
 	or dividing out the factor $(\xi^2 - k^2)$ (corresponding to 
the trivial  solution  $\xi^2 =k^2$) implies that $\xi$ at high frequency  satisfies  
\bal{0-9}
  \xi^2 &= k^2  - \frac{2in_0}{\sqrt{\xi k}}  f(0)  \big[ (k + \xi  ) e^{-i\xi b} 
+ i (\xi -k)e^{i\xi b}\big]e^{ikb}
 \nonumber \\ & \quad
+\frac{4in_0^2}{\xi k}  \big( f^2(0) - f^2(\pi)\big)e^{i2kb}. 
 	\end{align}

\section{Generalized Self Consistent Model}\label{sec2}
\

The Generalized Self Consistent Model (GSCM) developed by Yang and Mal  \cite{Yang94} was derived  using a self consistent scheme applied to the Waterman and Truell{'}s formula \cite{Waterman61}.  
Among the objectives of this section is to apply this scheme, which is very broad in scope, to the generalized Linton and Martin formula.
\

The basic idea follows  Christensen and Lo  \cite{Christensen79}. Instead of  considering  cylinders that are directly immersed in fluid, a ``three phase cylinder'' model is used which  assumes that each cylinder is surrounded by a cylindrical ring of fluid,  the whole being immersed in a outer region of equivalent fluid of unlimited extent. Hence, form functions   in this section correspond to three phase cylinders, and, as for cylinders, they may be expressed as a modal sum and calculated numerically \cite{Varadan80}.
\

Let $a$ be the radius of circular cylinders and $c$ ($0\leq c<1$) their concentration, the radius  $a_c$  of cylindrical rings is related to $a$ and $c$ by
\beq{jm1}
c=\frac{n_0 \pi a^{2}}{n_0 \pi a_c^{2}}=\frac{a^{2}}{a_c^{2}}.
\eeq
Let $\rho_{eff}$ and $k_{eff}$ be the effective properties of the equivalent outer fluid. The mass density $\rho_{eff}$ is  defined as the spatial average
\beq{jm2}
\rho_{eff}=c \rho_{cyl} +(1-c)\rho_{fluid},
\eeq
with $\rho_{cyl}$ and $\rho_{fluid}$ the cylinder and fluid mass densities. The wave number $k_{eff}$, which is unknown, is determined with the use of the self consistent scheme.
\

Without loss of generality, consider the Linton and Martin formula at the second order   in concentration. Let $\xi_{LM}$ be the Linton and Martin{'}s effective wave number  in the outer equivalent fluid, then we have 
\bal{jm3}
\xi_{LM}^{2}&=k_{eff}^{2}-4in_{0}f(k_{eff},0)+\frac{8 n_{0}^{2}}{ \pi k_{eff}^2}
\int\limits_{0}^{\pi}\dd\theta\, 
\nonumber \\ &   \quad \times
 \cot (\frac{\theta}{2}) \frac{\dd}{\dd\theta}[f(k_{eff},\theta) f(k_{eff},-\theta)],
 \end{align}
with
\beq{jm4}
f(k_{eff},\theta)= \sum\limits_n T_n(k_{eff})\e^{in \theta }
\eeq
and \cite{Varadan80} $n_{0}=c/ \pi a^{2}$. The self consistent scheme consists in assuming that $k_{eff}=\xi_{LM}$. From a physical point of view this means that the outer equivalent fluid is a medium in which the waves propagate in exactly the same manner as the coherent waves. Because $\xi_{LM}=k_{eff}$  there is no scattering due to the three phase cylinders in the outer equivalent fluid, and the medium can be considered as homogenized. It follows from \eqref{jm3} that $k_{eff}$ is given by the equation
\bal{jm5}
 f(k_{eff},0)&=\frac{2 n_{0}}{ i \pi k_{eff}^2}
\int_{0}^{\pi}\dd\theta\, \cot (\frac{\theta}{2}) 
\nonumber \\ &  \quad \times  
\frac{\dd}{\dd\theta}[f(k_{eff},\theta) f(k_{eff},-\theta)].
 \end{align}
It is worth mentioning that  at low concentration of cylinders the second term in Eq. \eqref{jm5} can be neglected, so that Eq. \eqref{jm5} reduces to
\beq{jm6}
f(k_{eff},0)=0, 
\eeq
which  corresponds to the equation for the CPA \cite{Sheng95}. In other words,  the CPA appears as the  approximation of Eq. \eqref{jm5} to first  order  in concentration.

Another way of presenting the self consistent scheme is to use an iterative procedure applied to Eq. \eqref{jm3}. Starting with $k_{0}=k$  we carry out the homogenization by employing Eq.\eqref{jm3}  to obtain $k_{1}$, and so on, so that 
\begin{align}
& 
k_{n+1}^{2}=k_{n}^{2}-4in_{0}f(k_{n},0)
\nonumber \\ & 
+\frac{8 n_{0}^{2}}{ \pi k_{n}^2}
\int_{0}^{\pi}\dd\theta\, \cot (\frac{\theta}{2}) \frac{\dd}{\dd\theta}[f(k_{n},\theta) f(k_{n},-\theta)]. \nonumber
\end{align}
The iteration is repeated until there is convergence,  $k_{n+1}\rightarrow k_{eff}$, the solution of \eqref{jm5}.  This procedure is of interest not only for computations but also for its physical interpretation. The effective wave number $k_{1}$ corresponds to a coherent wave that accounts for  the double interactions between cylinders according to the basic hypothesis of the Quasi Crystalline Approximation (QCA) \cite{Waterman61}. In the same way, $k_{2}$ involves double interactions in the medium characterized by $k_{1}$, the medium in which double interactions are already taken into account. So, relative to the fluid medium characterized by $k_{0}$, $k_{2}$ must deal with interactions of higher order, even if we do not know how to describe these explicitly.  In addition, the number of interactions between cylinders is  linked to the order of convergence of the scheme. Accordingly, it can be supposed that the use of the self consistent scheme applied to an explicit effective wavenumber, whatever it is, $k_{ISA}$, $k_{WT}$ or $k_{LM}$ (whatever the order in concentration), can improve the accuracy of the results while the concentration of cylinders increases. This is what  has been observed in  \cite{Yang94} when searching for dynamic effective mechanical properties of composites at low frequencies. Furthermore, just as the Linton and Martin approach is more efficient than the one of Waterman and Truell, Eq. \eqref{jm5} should be more efficient than the GSCM introduced by Yang and Mal, and also the CPA. Numerical calculations are beyond  the scope of this paper, but the different effective wavenumbers are compared at low frequency in the next section.

\section{The Rayleigh limit}\label{sec3}
\

In the Rayleigh or low-frequency limit, the size of the scatterers is assumed to be small  compared to the incident wavelength.  In this section  scatterers are specifically considered to be circular fluid cylinders of radius $a$, characterized by the density $\rho_{2}$ and the sound speed $c_{2}$. They are immersed in a fluid, characterized by the density $\rho_{1}$ and the sound speed $c_{1}$.  The Rayleigh limit then  corresponds to  $k_{1}a\rightarrow 0$ with $k_{1}=\omega/c_{1}$. It is then sufficient to take only the lowest order coefficients in the modal series \cite{Minonzio05}. More exactly, it can be shown that at low frequencies, only $T_{0}$ and $T_{\pm1}$ make a contribution, reducing the infinite matrix ${\bd Q}$ to a rank 3 matrix, see  \S\ref{sec1.5}. 
\

The goal of this section is to compare the  Rayleigh limit for the different effective wave numbers which correspond respectively to ISA, Waterman and Truell, Linton and Martin, CPA, the GSCM developed by Yang and Mal \cite{Yang94}, noted here G-WT, and the GSCM introduced in \S\ref{sec2} from the Linton and Martin approach, noted here G-LM.
\

\subsection{Effective wave numbers}
We first calculate  effective wave numbers using  the Rayleigh limit of the far-field scattering function associated with  cylinders which are directly immersed in fluid:  
\begin{subequations}\label{jm8}
\bal{jm8a}
T_{0}&=\frac{i\pi }{4}  \bigg(\frac{\rho_{1}c_{1}^{2}}{\rho_{2}c_{2}^{2}}-1\bigg)
(k_{1}a)^2,  
 \\
T_{\pm1}&=\frac{i\pi }{4}    \bigg(\frac{\rho_{2}-\rho_{1}}{\rho_{2}+\rho_{1}}\bigg)
(k_{1}a)^2.
\end{align}
\end{subequations}
These coefficients can be derived from those corresponding to circular elastic cylinders immersed in fluid \cite{Minonzio05}, if the shear and longitudinal velocities are equal to $0$ and $c_{2}$ respectively (Scattering coefficients are denoted by $R$ in Ref. \cite{Minonzio05} with $R_{0}=T_{0}$ and $R_{\pm1}=-T_{\pm1}$). 

The use of Eqs.\eqref{jm8} leads to the following results for the ISA, Waterman and Truell (WT) and Linton and Martin (LM). First
\bal{jm10}
\big(\frac{k_{eff}}{k_{1}}\big)^{2} &= 1+\bigg( \frac{\rho_{1} c_{1}^{2} }{\rho_{2} c_{2}^{2}}+\frac{\rho_{2}-3\rho_{1}}{\rho_{1}+\rho_{2}}\bigg) c 
 \nonumber \\ & 
 \equiv \big(\frac{k_{eff}}{k_{1}}\big)^{2}_{ISA} \quad \text{(ISA)}, 
\end{align}
in terms of which the other two are 
\bal{jm11}
 \big (\frac{k_{eff}}{k_{1}}\big)^{2} &=
 \big(\frac{k_{eff}}{k_{1}}\big)^{2}_{ISA}
 + 2  c^{2}\bigg( \frac{\rho_{2}-\rho_{1}}{\rho_{2}+\rho_{1}}\bigg)
 \nonumber \\ & \quad
 \times
 \begin{cases}
 \bigg(\frac{\rho_{1} c_{1}^{2} }{\rho_{2} c_{2}^{2}}-1
\bigg)   & \text{(WT)}, 
\\
 \bigg(\frac{\rho_{1} c_{1}^{2} }{\rho_{2} c_{2}^{2}}-
\frac{2\rho_{1}}{\rho_{2}+\rho_{1}} \bigg) 
& \text{(LM)}  .
\end{cases}
\end{align}
Note that the latter two  effective wavenumbers are almost the same if the densities $\rho_{1}$ and $\rho_{2}$ are close to one another in value. However,  even at low frequency  where cylinders look like ``point scatterers'', the Waterman and Truell and the Linton and Martin approaches give different results as soon as $\rho_{2}$ is not close to $\rho_{1}$.
\

\subsection{Wave numbers from  the self consistent scheme}
We now consider  effective wave numbers  obtained by the self consistent scheme, which uses  scattering coefficients for  the ``three phase cylinder'' as described in \S\ref{sec2}. The coefficients are calculated as outlined in \cite{Veksler93}, with the results 
\bal{jm9}
&T_{0}=\frac{i \pi }{4}\bigg(
(1-c)\frac{\rho_{eff}}{\rho_{1}}
+c\frac{\rho_{eff} c_{1}^{2}}{\rho_{2} c_{2}^{2}} 
-\frac{k_{eff}^2}{k_{1}^2} \bigg)(k_1 a_c)^2,
\nonumber \\
&T_{\pm1} =\frac{i\pi  }{4 }    \, 
F(\rho_{eff}) (k_{eff} a_c)^2,
\end{align}
where  $\rho_{eff} =  (1-c)\rho_1 +c\rho_2 $ from Eq. \eqref{jm2}, and
\begin{align}
&\qquad F(\rho_{eff})= 
\nonumber \\ &
\frac{(1-c)(\rho_{1}^{2}-\rho_{2} \rho_{eff})+(1+c)\rho_{1}(\rho_{2}-\rho_{eff})}
{(1-c)(\rho_{1}^{2}+\rho_{2} \rho_{eff})+(1+c)\rho_{1}(\rho_{2}+\rho_{eff})}.
\nonumber 
\end{align}
Note that  Eqs. \eqref{jm8} follow from Eqs. \eqref{jm9}  formally if  we first put    $k_{eff}=k_{1}$ and $\rho_{eff}=\rho_{1}$, in order to identify the outer equivalent fluid with the fluid itself, and then cancel the ring of fluid by setting $a_c=a$, which implies  $c=1$.
\

We now consider three approaches based on self consistent schemes: the CPA, the generalized self consistent method based on  Waterman and Truell (G-WT), and the same for Linton and Martin's approach (G-LM).  We find 
\bal{jm13}
& \big( \frac{k_{eff}}{k_{1}}\big)^{2} =\frac{\rho_{eff}}{\rho_{1}}\big[1+
\big(\frac{\rho_{1} c_{1}^{2} }{\rho_{2} c_{2}^{2}}-1 \big)c \big]
\nonumber \\ & \quad 
 \times
\begin{cases}
\big[1-2F(\rho_{eff}) \big]^{-1} & \text{(CPA)} , 
\\
\big[1+2F(\rho_{eff}) \big] & \text{(G-WT)} , 
\\
\bigg[\frac{ 1+2F(\rho_{eff})}{
 1-2F^{2}(\rho_{eff})  } \bigg] & \text{(G-LM)} .
\end{cases}
\end{align}
In order to compare these effective wave numbers with those of eqs. \eqref{jm10} and \eqref{jm11} in the previous subsection,  we have to perform an asymptotic expansion with regard to the concentration of scatterers. It follows that
\beq{jm17}
F(\rho_{eff})= \frac{(\rho_{1}-\rho_{2})^{2}}{2\rho_{1}}
\bigg\{
\frac{-c}{\rho_{1}+\rho_{2} } 
+\frac{c^{2}}{2 \rho_{1} } +...\bigg\} ,
\nonumber 
\eeq
and we find at the second order  in concentration
\bal{jm18}
&\big (\frac{k_{eff}}{k_{1}}\big)^{2} =\big (\frac{k_{eff}}{k_{1}}\big)^{2}_{ISA}+
2 c^2 \bigg(\frac{\rho_{2}-\rho_{1}}{\rho_{2}+\rho_{1}} \bigg)
\nonumber \\ & 
\times
\begin{cases}
\big[\frac{\rho_{1} c_{1}^{2} }{\rho_{2} c_{2}^{2}}-
\frac{2\rho_{1}}{\rho_{2}+\rho_{1}} -\frac{(\rho_{1}-\rho_{2})^{2}}{4\rho_{1} (\rho_{1}+\rho_{2})}\big]  
& \text{(CPA)} , 
\\ \big[\frac{\rho_{1} c_{1}^{2} }{\rho_{2} c_{2}^{2}}-
\frac{2\rho_{1}}{\rho_{2}+\rho_{1}} +\frac{(\rho_{1}-\rho_{2})^{2}}{4\rho_{1} (\rho_{1}+\rho_{2})}\big]
& \text{(G-WT)} , 
\\
\big[\frac{\rho_{1} c_{1}^{2} }{\rho_{2} c_{2}^{2}}-
\frac{2\rho_{1}}{\rho_{2}+\rho_{1}} \big]  & \text{(G-LM)} .
\end{cases}
\end{align}

As expected, all three methods give the same result at the first order in concentration.  
At the second order, the CPA and G-WT introduce the same additional term as compared to  Linton and Martin{'}s approach, but with the opposite sign in each.   It is significant that 
the self consistent scheme applied to  Linton and Martin{'}s formula does not modify the result, at least at this order.   The same cannot be said of the CPA and the Waterman and Truell methods. Thus, we may conclude that   Linton and Martin{'}s approach and the G-LM can be considered  ``self consistent'' methods.  Finally, we note  that the additional terms in the CPA and in the 
G-WT in Eq.  \eqref{jm18} are very small if the densities $\rho_{1}$ and $\rho_{2}$ are close in value. Hence,  all the methods are equivalent at low frequency when the densities $\rho_{1}$ and $\rho_{2}$ are equal. Of course, these results say nothing about what happens at higher frequency and at higher concentration.

\section{Conclusions}
Implications of the quasi-crystalline approximation (QCA) on the effective wave number beyond the dilute concentration limit have been described.  Equation \eqref{036} is  the starting point for all further developments, and as such represents the fundamental  result of the paper.  It splits the implicit form  of the effective wavenumber into two distinct parts, one defined  by the single scatter T-matrix,  
${\bd T}$, and the other by the spatial arrangement of the scatterers, $\bar{\bd Q}$.  In this paper we have used to the hole correction, for which  $\bar{\bd Q}$ is given by \eqref{3e}.    More generally, this matrix has elements 
\beq{0223}
\bar{Q}_{mn}     = 
\frac{L_{m-n}  }{-i4n_0}  e^{-i\xi x_1}
+ \frac{ e^{i(k-\xi) x_1}} {   2k( \xi -k)} 
- \frac{ 1} {     \xi^2 -k^2}  
,
\eeq
where $L_{m-n} $ is defined by Eq. \eqref{-55} for arbitrary pair correlation function.  The QCA is exact for a regular array of scatterers, in which case  $\bar{Q}_{mn}$ can be reduced to a known lattice sum.  Equation \eqref{036} therefore provides a   formula for determining the dispersion curves of a regular array.   This and other implications will be examined elsewhere.  


\end{document}